      \documentclass{elsarticle}
      \usepackage{hyperref}
      
            \hypersetup{colorlinks,
	      citecolor=blue, 
	      filecolor=black,
	      linkcolor=black,
	      urlcolor=blue   
          }
          
      \journal{Journal of Food Structure}
      \usepackage{graphics}
      \usepackage[utf8x]{inputenc}		
      \usepackage{amsmath}                     
      \usepackage{amssymb}                     
      \usepackage{multirow}			

      \newcommand{\angstrom}{\textup{\AA}}

      \DeclareFontFamily{U}{euc}{}
      \DeclareFontShape{U}{euc}{m}{n}{<-6>eurm5<6-8>eurm7<8->eurm10}{}%
      \DeclareSymbolFont{AMSc}{U}{euc}{m}{n} 
      \DeclareMathSymbol{\umu}{\mathord}{AMSc}{"16} 

      \newcommand{\umum}{\hspace{0.6 mm}\,\umu\textrm{m}}

\begin{document}
      \begin{frontmatter}
      

\title{
      Ptychographic X-ray computed tomography of extended colloidal networks in food emulsions
      }

\author[mymainaddress]{Mikkel Schou Nielsen \corref{mycorrespondingauthor}}
\cortext[mycorrespondingauthor]{Corresponding author}
\ead{schou@nbi.ku.dk}

\author[mysecondaryaddress]{Merete B\o gelund Munk}

\author[mythirdaddress]{Ana Diaz}

\author[myfourthaddress]{Emil B\o je Lind Pedersen}

\author[mythirdaddress]{Mirko Holler}

\author[myfifthaddress]{Stefan Bruns}

\author[mysecondaryaddress]{Jens Risbo}

\author[mymainaddress]{Kell Mortensen}

\author[mymainaddress]{Robert Feidenhans'l}

\address[mymainaddress]{Niels Bohr Institute, University of Copenhagen, Universitetsparken 5, 2100 Copenhagen, Denmark}
\address[mysecondaryaddress]{FOOD, University of Copenhagen, Rolighedsvej 30, 2000 Frederiksberg, Denmark}
\address[mythirdaddress]{Paul Scherrer Institute, 5232 Villigen – PSI, Switzerland}
\address[myfourthaddress]{Department of Energy Conversion, Technical University of Denmark, Frederiksborgvej 399, 4000 Roskilde, Denmark}
\address[myfifthaddress]{Department of Chemistry, University of Copenhagen, Universitetsparken 5, 2100 Copenhagen, Denmark}

    \begin{abstract}
     As a main structural level in colloidal food materials, extended colloidal networks are important for texture and rheology. By obtaining the 3D microstructure of the network, macroscopic mechanical properties of the material can be inferred. However, this approach is hampered by the lack of suitable non-destructive 3D imaging techniques with submicron resolution. 

     We present results of quantitative ptychographic X-ray computed tomography applied to a palm kernel oil based oil-in-water emulsion. The measurements were carried out at ambient pressure and temperature. The 3D structure of the extended colloidal network of fat globules was obtained with a resolution of around 300 nm. Through image analysis of the network structure, the fat globule size distribution was computed and compared to previous findings. In further support, the reconstructed electron density values were within 4\% of reference values.
    \end{abstract}



      
      \end{frontmatter}


\section{Introduction}

Extended colloidal networks constitute one of the main structural elements in multi-phase food materials such as butter, chocolate, cream cheese, whipped creams, ice cream, cheese and yogurt. Focus has been directed toward establishing the relationship between the structure of such networks on one side, and the macroscopic mechanical properties and sensorial textural properties on the other side \citep{narine_review_1999, Heertje2014}. On a qualitative level, correlations have been established between microscopic observations of loose networks and low modulus or relationship between a more dense network and a higher modulus (see for example \citep{kaufmann_cooling_2012,wiking_relations_2009,buldo_butter_2011}). On a more quantitative level, a fractal description has been employed for inorganic colloidal networks \citep{shih_scaling_1990}, for protein systems \citep{stading_microstructure_1993,vreeker_fractal_1992} and fat systems \citep{marangoni_1998,narine_physreve_1999,marangoni_edible_2012} leading to a quantitative relationship between structure and mechanical properties. 

The experimental study of such networks has traditionally been based on light microscopy such as polarized light microscopy (PLS) and confocal laser scanning microscopy (CLSM). From the microscopy measurements, the network properties have for the most been analyzed using 2D slices. An extension to 3D imaging can be done by forming 3D image stacks from consecutive micrography at different depths in the food product. In this way, Litwinenko used transmitted PLS and image deconvolution to study the fractal properties of a fat crystal network in two and three dimensions \citep{litwinenko_3dfat_2004}.
Similarly, CLSM can be applied as an 3D imaging method within food science. However, the obtained stacks are limited by the optical system and the laser penetration depth to some tens of microns from the top surface \citep{Durrenberger2001}. Furthermore, although a sub-micron spatial resolution is possible, scattering in the sample may limit the resolution along the vertical axis to the micron range \citep{Fredrich1999}. In addition, the need of staining in CLSM may introduce artifacts and limit the in-situ applicability \citep{Durrenberger2001}. 
To our knowledge CLSM has not been used to image the 3D structure of colloidal networks in food materials.

In recent years, X-ray phase-contrast computed tomography (CT) has emerged at synchrotron facilities as a non-destructive 3D imaging modality, and has been successfully applied to study the microstructure of a range of food products \citep{FalconePhaseBread2004, verbovenPomeFruit2008,JensenPorc2011,MiklosHeatMeat2015}.
One of the most recent techniques for obtaining the X-ray phase-contrast modality is ptychographic X-ray computed tomography (PXCT) \citep{dierolf2010ptychographic}. PXCT is a 3D nano-imaging technique that offers a spatial resolution in the 100 nm range. Unlike traditional X-ray microscopy, the spatial resolution is not dependent on objective lenses. Instead, spatial information is retrieved from the recorded diffraction of a coherent X-ray beam, and the spatial resolution is only limited by the angular spread of the scattered intensity. 
In addition, PXCT provides quantitative information by reconstructing the full 3D electron density distribution of the specimen \citep{DiazQuant2012,Thibault2014}. The technique is well-suited for in-situ measurements as it allows for sample environments at room temperature and ambient pressure. Previously, it has successfully been applied for an in-situ study of water uptake in a single silk fiber \citep{ptycho_silk_2013}. 
Altogether, PXCT is a promising candidate for 3D imaging of extended networks in food products.

As a model system for studying extended colloidal networks, a palm kernel oil (PKO) based oil-in-water emulsion is presented. These PKO emulsions are used as whippable creams for decorations of cakes where the fat globule network formation is important. 
Design of emulsions for whipping relies on tuning the propensity of partial coalescence of the oil droplets. Initially PKO emulsions are normally liquid, and first upon whipping the material is transformed to a foam of rather high viscosity and stability. However, too high propensity for partial coalescence can lead to product flaws such as solidification of the liquid emulsion to solid pastes during transport.  
As an example of a product flaw, a PKO emulsion exhibited pre-whipping solidification upon addition of two different combinations of lactic acid ester of monoglyceride (LACTEM) and unsaturated monoglyceride (GMU) \citep{munk_whippable_2013, munk_coalescence_2015}. In these two systems, 2D CLSM micrographs of the lipid phase revealed large irregular aggregates and formation of extended networks of fat globules. In addition, increased hardness and viscoelastic modulus were observed. The added emulsifiers are believed to induce partial coalescence of the fat globules and transform the emulsion spontaneously from liquid to semi-solid \citep{munk_coalescence_2015}. However, due to strong multiple scattering of the laser light, the 3D structure of the network could not be determined using CLSM.

Thus, the exact extend and composition of the network of fat globules in 3D are still unknown. In addition, exactly how the water and lipid phases are located remain to be directly observed. In this study, a PKO emulsion with two combinations of LACTEM and GMU emulsifiers are measured with PXCT and compared to 2D CLSM micrographs. The 3D structure of both water and lipid phase as well as the quantitative electron density values are investigated. 

\section{The X-ray phase-contrast modality}

The type of image contrast acquired in X-ray tomograms depends on the interaction between the X-rays and matter. Both refraction and absorption of X-rays in matter are given by the full complex index of refraction \citep{als}

\begin{align}
 n(\mathbf{r}) &= 1-\delta(\mathbf{r}) + i\beta(\mathbf{r}),
\end{align}

where the real part $\delta$ accounts for the refraction and the imaginary part $\beta$ 
for the absorption. 
In X-ray phase-contrast techniques such as PXCT, the 3D distribution of $\delta(\mathbf{r})$ 
is reconstructed in the tomogram. Thus, the gray levels in the 
resulting images are due to the spatial variations of $\delta(\mathbf{r})$ in the material. These can be related to the electron density $\rho_e(\mathbf{r})$ as 

\begin{align}\label{eq:beta:delta}
\rho_e(\mathbf{r}) &= \frac{2\pi \delta(\mathbf{r})}{r_0\lambda^2},
\end{align}\ 

where $r_0$ is the Thomson scattering length and $\lambda$ the X-ray wavelength. For materials with known atomic composition, values of $\rho_e$ obtained by PXCT can be compared to calculated values. For mixtures of materials of different weight-percentages $\textrm{w}_j$, the total electron density can be calculated as


\begin{align}\label{eq:rhoe}
\rho_e&=N_A\rho_m\sum_j \textrm{w}_j \frac{Z_j}{M_j}
\end{align}

where $N_A$ is Avogadro's constant, ${\rho_m}$ is the mass density and $Z_j$ and $M_j$ are the number of electrons and the molar mass of the jth material, respectively. 

\section{Material and methods}

\subsection{Materials}

Emulsifiers were of commercial food grade and all were provided by Palsgaard A/S (Juelsminde, Denmark): Lactic acid ester of monoglyceride (LACTEM) made from fully hydrogenated palm oil and rape-seed oil (C16:0 and C18:0 $>97\%$ of fatty acids); unsaturated monoglycerides (GMU) made from sunflower oil (C18:1 $>81\%$). The stabilizer mixture 
(Palsgaard A/S, Juelsminde, Denmark) contained microcrystalline cellulose (MCC), sodium carboxymethylcellulose (CMC) and disodium phosphate. Hydrogenated palm kernel oil (PKO) was obtained from AAK (Karlshamn, Sweden), sodium caseinate from DMV International (Veghel, The Netherlands) and sugar from Nordic Sugar (Nakskov, Denmark). 
Fatty acid composition of the PKO has previously been determined \citep{munk_coalescence_2015}.
    
\subsubsection{Emulsion blend preparation}

Sodium caseinate (0.6 wt.\%), stabilizer mixture (0.6 wt.\%) and sugar (10 wt.\%) were dispersed in water under continuous stirring and put aside for 4 h to hydrate proteins. Melted PKO (25 wt.\%), either LACTEM (0.55 wt.\%) alone or with GMU (0.15 wt.\%) were mixed with the water phase, and the mixture was heated to 80 $^\circ$C. A pre-emulsion was obtained by mixing with a high-shear blender (Ultra-Turrax, IKA, NC, USA) for approximately 20 s. Homogenization was subsequently carried out on a two-stage high-pressure valve homogenizer (PandaPlus 2000. GEA Niro Soavi, Parma, Italy) at 150/50 bar followed by cooling in a turbular heat exchanger to 30 $^\circ$C. Immediately afterwards, small samples of the emulsions were prepared in micropipettes using a syringe. Subsequently, all emulsion samples were stored at 5 $^\circ$C.

\subsubsection{Micropipette preparation}
    
The micropipettes were prepared from thin wall borosilicate capillaries with filament (Harvard Apparatus UK, Cambridge, UK) by using a micropipette puller (Sutter Instrument, CA, USA). The capillaries were pulled from the original 0.94 mm to an inner diameter at the tip in the range from 15 to 20$\umum$. Prior to measurements, the micropipettes were mounted on custom-made tips.

\subsection{Confocal laser scanning microscopy}
    
Micrographs of the lipid phase in the emulsion were obtained by an inverse confocal laser scanning microscope (Leica TCS SP5, Heidelberg, Germany). Total concentration of 1 ppm BODIPY 493/503 (4,4-difluoro-1,3,5,7,8-pentamethyl-4-bora-3a,4a-diaza-s-indacene) (Invitrogen, Carlsbad, USA) dissolved in dimethyl sulfoxide was added to emulsions to stain lipids. The excitation wavelength was 488 nm and the emission bandwidth was 500–570 nm. A water immersion objective (HCX PL APO lambda blue 63.0 $\times$ 1.20 water UV) was used, and the image size was set to 1024 $\times$ 1024 pixels. A thin layer of stained emulsion was applied on a large standard cover glass and covered by another
cover glass to avoid evaporation/drying. Around 20 micrographs were acquired of each emulsion preparation.

\subsection{Ptychographic X-ray computed tomography}
     
\begin{figure}[tb]
  \centering
  \includegraphics[width=0.9\textwidth]{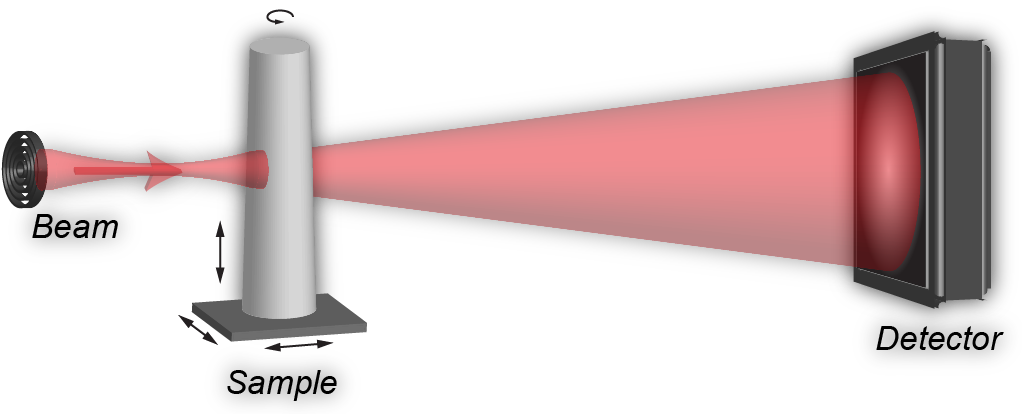}
  \caption{ \label{fig_setup}
  \footnotesize{A sketch of the PXCT setup. The beam is shaped by a Freznel zone-plate before illuminating the sample. At the detector position, a diffraction pattern of the illumination is acquired. For the ptychographic scans, the sample may be translated in all directions. Rotation around the vertical axis is used for the tomography measurements.}
  }
  \end{figure}

The PXCT measurements were conducted at the cSAXS beamline at the Swiss Light Source, Paul Scherrer Institut, Switzerland, using 6.2 keV X-rays and the setup described in \citet{holler2014}. A sketch of the setup is shown in figure \ref{fig_setup}

A coherently illuminated freznel zone-plate of diameter $170\umum$ and 60 nm outer-most zone width was used to shape the illumination onto the sample. The sample was placed 1.8 mm downstreams of the focal point where the beam had a diameter of about $6\umum$. Because the initial illumination was causing clear radiation damage effects on the emulsion samples, the flux was reduced by a factor of 10 by opening the undulator gap by about $30\umum$ from the maximum intensity. The final flux of about $1\times10^8$ photons/s did not cause any visible damage to the emulsions.

Ptychographic scans were performed over different fields of views (FoV) for each sample, as seen in table \ref{tab_ptycho}. The scanning pattern followed a Fermat spiral trajectory \citep{HuangFermat14} with an average spacing between points of $2\umum$. 
A diffraction pattern was recorded at each point  with an exposure time of 0.1 s with an Eiger 500k detector as in \citet{DinapoliEiger2011, Guizar-Sicairos_eiger_2014} which was located 7225 mm downstream of the sample position behind a He-filled flight tube. Ptychographic scans were repeated at several angular positions of the specimen with respect to the X-ray beam, covering a 180$^\circ$ range. The total number of diffraction patterns and angular projections were different for each sample as detailed in table \ref{tab_ptycho}. The total measurement time varied between 3-10 hours.

Four tomograms were acquired in total. Two were of the emulsion containing both LACTEM and GMU and two of the emulsion containing only LACTEM as indicated in table \ref{tab_ptycho}. The latter two tomograms were conducted on the same micropipette.

\begin{table}[h!tb]
\centering
\renewcommand{\tabcolsep}{2.5mm}
\renewcommand{\arraystretch}{1.2}
\begin{tabular}{l|c|c|c|c}
\multicolumn{5}{c}{\textbf{Measurement details}}\\
\hline
 & Sample 1 & Sample 2 & Sample 3 &  Sample 4 \\
\hline
Emulsifier mix. & L+G & L+G & L & L \\
FoV [$\umum^2$] & 48x40 & 30x20 & 30x20 & 45x27 \\
No of dif. pat. & 481 & 151 & 170 & 311 \\
Angular projs & 200 & 250 & 300 & 400\\
\hline
Spatial res. [nm] & 342 & 207 & 280 & 355\\
Total dose [MGy]	  & 1.0 & 1.2 & 1.5 & 2.0\\ 
\hline
\end{tabular}
\caption{\label{tab_ptycho}Information on sample composition, settings for the PXCT measurements, and the spatial resolution in the reconstructed tomograms. L=LACTEM and G=GMU. }
\end{table}

From the acquired data, first 2D ptychographic projections of the phase were reconstructed using a combination of the difference map algorithm \citep{ThibaultProbe2009} and a maximum likelihood optimization \citep{Guizar-SicairosPhaseRetriveal2008} used as a refinement step \citep{thibaultMlRefinement2012}. 
For ptychographic reconstructions 
In order to aim at a final 3D tomographic spatial resolution of 100 nm, an area on the detector of 400x400 pixels around the direct beam was selected such that the reconstructed ptychographic projections had a pixel size of $49\times49 \,\mathrm{nm}^2$. Note that the pixel size and spatial resolution are not the same. Multiple pixels may be required to resolve features in the final images. The reconstructed phase projections for each tomogram were further processed and combined in a tomographic reconstruction using the filtered back projection as described in \citet{Guizar-Sicairos_ptychomethod2011}. The tomographic slices were corrected for a constant bias off-set by forcing the mean value of the air outside the sample to be zero. The processing was done using a software implementation in MATLAB at the cSAXS beamline.

The spatial resolution of the final tomograms was evaluated by comparing their Fourier shell correlation (FSC) curves with the 1/2-bit threshold curve \citep{van2005fsc}. The sample stability turned out to be the limiting factor. Whereas the 3D resolution in a region with the micropipette was around 100 nm for all samples, the resolution in a region of the emulsions themselves were between 207-355 nm, as seen in table \ref{tab_ptycho}. This reduction can be ascribed to sample movement during measurements and could induce a blurring of the features in the final tomograms.

The dose $d$ imparted on the specimen was estimated using the expression $d=N_0\times E\times\mu/\rho_m$ where $\mu/\rho_m$ is the mass attenuation coefficient, $N_0$ is the number of incident photons per unit area, $E$ is the photon energy, and $\rho_m$ is the mass density \citep{Howells2009}. Since the illumination, average step size and exposure time was kept constant, the dose was the same for all 2D ptychographic scans. Using $N_0=2.8\times10^6 \, \textrm{photons}/\umum^2$ and $\mu/\rho_m=18 \,\textrm{cm}^{2}/\textrm{g}$ for the emulsion at 6.2 keV \citep{nist_xray_data}, the dose was $d = 5.0\times 10^{3}\, \textrm{Gy}$ per ptychographic scan. The total dose on each sample depended only on the number of angular projections, as shown in table \ref{tab_ptycho}. 


\subsection{Image analysis}

To perform noise correction on the tomographic slices, a customized iterative 3D implementation of the non-local means algorithm \citep{Buades2005} was applied. The level of noise and artifacts in the PXCT were assumed to be independent of the signal intensity. The denoising algorithm was implemented using Python.

An alpha-level Markov random field (MRF) segmentation as described in \citet{pedersen2015improving}
was applied for segmentation of the tomograms into air, micropipette, water, lipid and cellulose phases. First, the data was modeled as a mixture of distribution functions by assigning a probability distribution for each ascribed phase in the tomogram. Due to the sample movement during measurements, the water and lipid phases appeared blurred in the tomograms, and a model of two Gaussian distributions was used for these (see supporting material).
After assigning probability distributions, the spatial information of the data was incorporated into the segmentation process by modeling the data as an isotropic MRF \citep{li_2009}. The MRF smoothing parameter was set to 0.5. To find the optimal segmentation solution the multi-labeling problem was solved using graph cuts with alpha expansions as described in \citet{boykov_graphcut_2001}. 
Image analysis as well as visualization of the tomograms was performed using own software implemented in MATLAB.

\subsubsection{Quantitative parameters}

From the segmented tomograms, the percent object volume (POV) values for the identified lipid, water and stabilizer phases were calculated. Reference POV values were obtained by calculation based on the emulsion blend preparation. The continuous phases and isolated regions of the phases were obtained by applying a connected components labeling algorithm. The volume fraction of the total lipid phase found in the largest connected network was calculated.

Assuming that the lipid phase in the emulsion had only partially coalesced, the individual fat globules would still be visible in the 3D network. To investigate this, the structure of the lipid phase was divided into smaller local domains using a watershed approach. From the segmented lipid phase, a distance map from the lipid to the surroundings was calculated before the watershed algorithm was applied. From the volume of each identified lipid domain, the equivalent diameter was extracted assuming a spherical shape. In addition, the mean diameter and the size distribution of these diameters were calculated.

A mean electron density was calculated for both the emulsion phases as well as the micropipette. The segmented phase was used as a mask on the acquired tomogram to identify the electron densities for each phase. Reference values were obtained from the known chemical composition and mass density of the components in the phases using equation \eqref{eq:rhoe} (see supporting material). The reference lipid phase value was calculated from the fatty-acid composition of the partly hydrogenated PKO which has been reported in \citet{munk_coalescence_2015}. For the micropipette, the atomic composition of the borosilicate material was supplied by the manufacturer \citep{HarvardCapillary}.

\section{Results}

\begin{figure}[tb]
  \centering
  \includegraphics[width=0.98\textwidth]{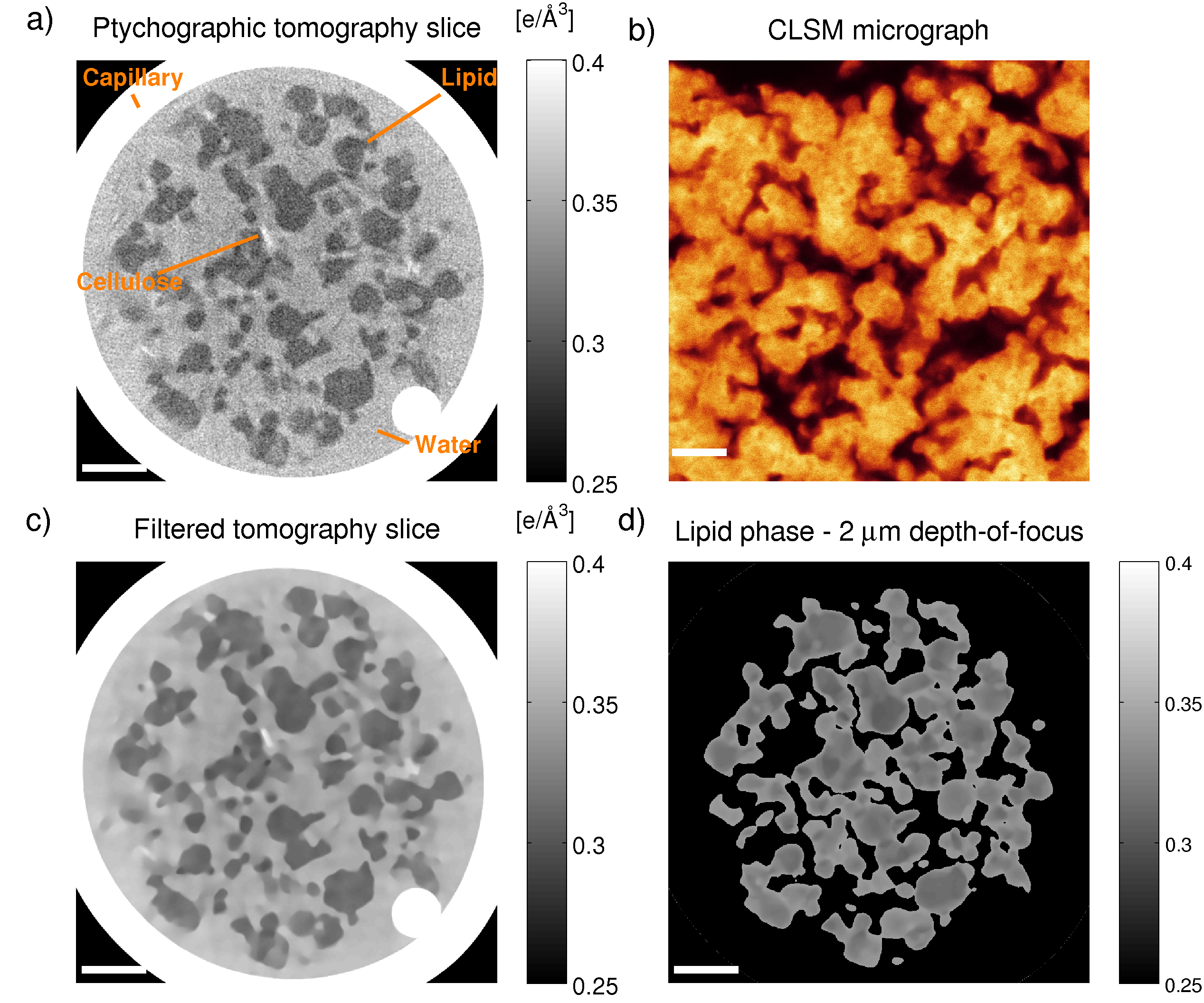}
  \caption{ \label{fig_slices}
  \footnotesize{A ptychographic tomographic slice (panels a and c) of the sample 1 emulsion compared to a CLSM micrograph (panel b). The tomographic slice before a) and after c) noise correction depicts the electron density of the sample in $[e/\textrm{\AA}^3]$. In panel d) shows only the lipid phase of the tomographic slice where the depth-of-focus has been increased to $2\umum$ by averaging over 40 slices in the z-direction. The length of the scale bar corresponds to $10\umum$.}
  }
  \end{figure}

In figure \ref{fig_slices}, a tomographic slice from the PXCT measurement of sample 1 is displayed before and after noise correction in panels a) and c), respectively. As the lipid phase has a lower mass density, and thereby lower electron density, than the water phase, it appears dark gray while the water phase appears light gray. Due to movement during the tomographic measurements, the lipid phase will be distributed over a larger volume. This causes a smearing in the reconstructed images which can be seen clearly in panel c) as brighter regions surrounding the lipid phase.

In panel b) a CLSM micrograph of the stained lipid phase is shown of an emulsion from the same batch as sample 1. When comparing the micrograph with the tomographic slice in panel c), some differences of the morphology of the lipid phase are seen. While the lipid phase in the micrograph forms large irregular aggregates, the tomographic slice shows individual fat globules or smaller aggregates. 

Part of this discrepancy can be explained by the difference in vertical spatial resolution of the two methods. While the PXCT  has an effective resolution of 200-300 nm with a slice thickness of 49 nm, the vertical resolution of the micrograph is in the micron range. Thus, more lipid phase will be visible in the micrograph than in a single tomographic slice. For a better comparison of the two modalities, the mean lipid phase from a stack of 41 tomographic slices with a total thickness of $2\umum$ is depicted in panel d). The segmented tomogram has been used as a mask to identify the lipid phase. As in the CLSM micrograph, the lipid phase in panel d) is seen to consist of large irregular aggregates with a similar morphology as in panel b).

    
      \begin{figure}[tb]
      \centering
      \includegraphics[width=0.98\textwidth]{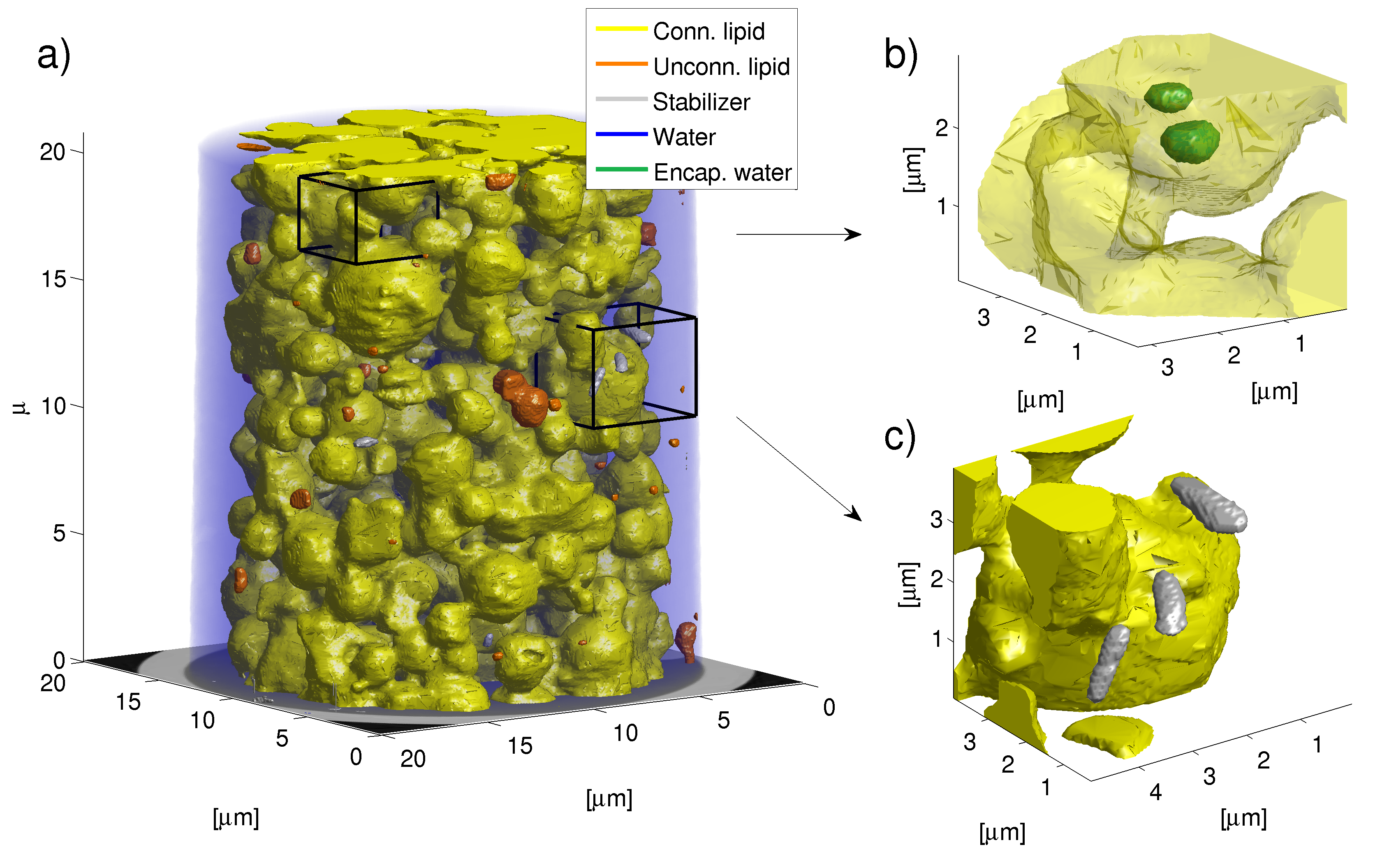}
       \caption{ \label{fig_3dsample2}
      \footnotesize{3D representation of the sample 2 emulsion. a) The full emulsion with the water phase shown in blue, the stabilizer in phase in gray, the connected lipid phase in yellow and isolated lipid regions in orange. b) A close-up of the sub-volume in the top box of a) depicting an isolated water pocket inside the connected lipid phase. c) A close-up of the stabilizer phase in the sub-volume of the bottom box of panel a).}
      }
      \end{figure}

A 3D representation of the segmented tomogram of sample 2 is illustrated in figure \ref{fig_3dsample2} (See supporting material for sample 1, 3 and 4). Displayed in panel a) are the identified emulsion phases; lipid, water and stabilizer. No air pockets were found in the volume inside the micropipette in this nor any of the other tomograms. 
The largest connected part of the lipid phase is shown in yellow while the remaining isolated regions are shown in orange. As seen, the vast majority of the lipid phase was connected in a single network which was also observed for the other phases. This space-filling lipid network constituted more than 98\% of the total lipid volume in all samples. 
The surrounding water phase is shown in blue in panel a) of figure \ref{fig_3dsample2}. Even though more than 99\% of the water volume was situated in this continuous phase, isolated water pockets inside the lipid phase were observed as illustrated in green color in the close-up in panel b).
Besides the lipid and water phases, a third stabilizer phase was identified which is depicted in gray in panels a) and c). From the elongated shape and the location in the water phase, these regions were interpreted as microcrystalline cellulose from the stabilizer mixture.

      \begin{figure}[tb]
      \centering
      \includegraphics[width=0.98\textwidth]{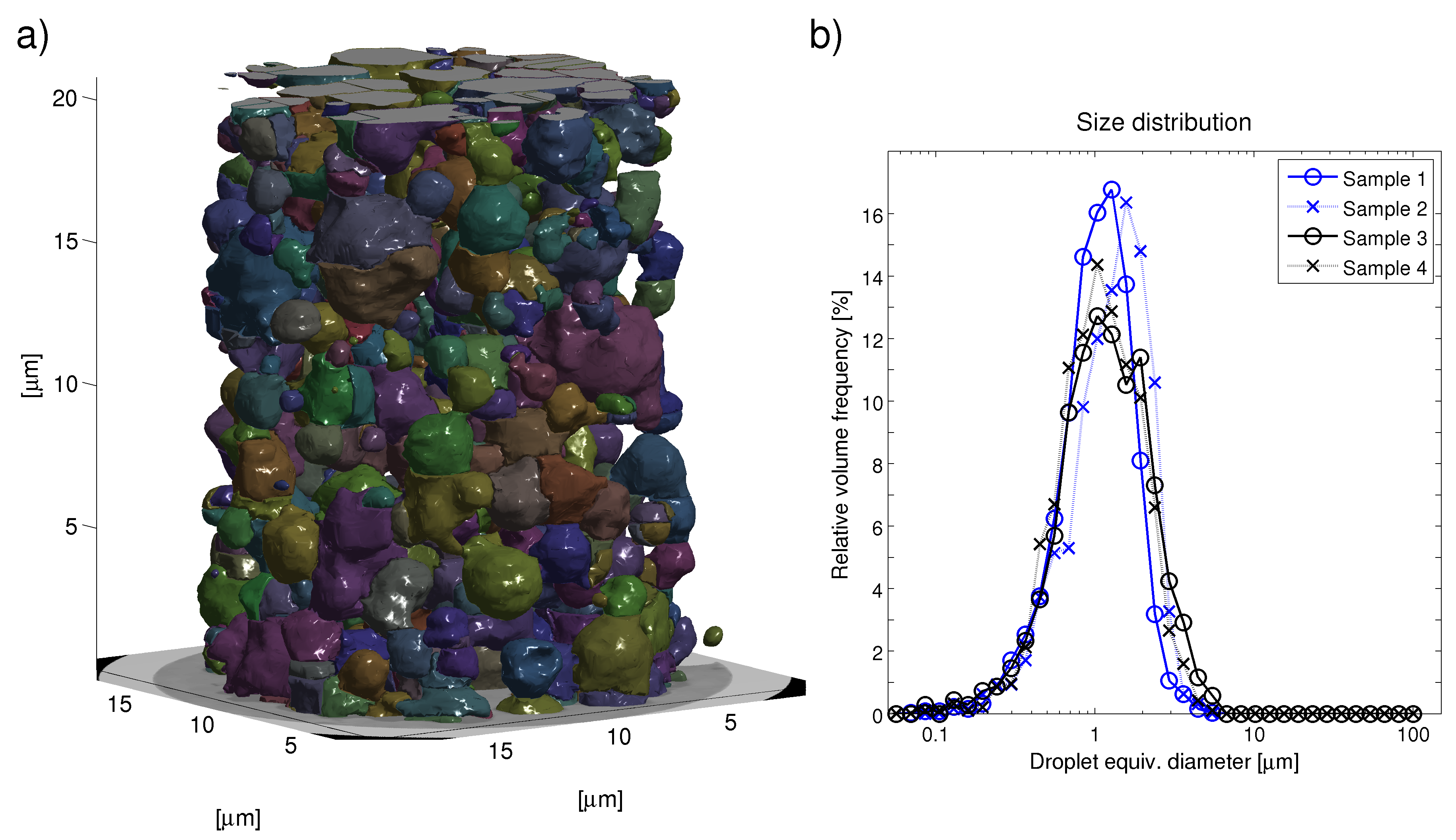}
       \vspace{0.5ex}
       \caption{ \label{fig_psd}
      \footnotesize{ a) A 3D representation of the resulting labelmap after applying the watershed algorithm to the connected lipid phase from figure \ref{fig_3dsample2} panel a). The coloring distinguishes individual domains. b) The size distributions of the equivalent diameters for all four samples. }
      }
      \end{figure}

The morphology of the lipid network in figure \ref{fig_3dsample2} panel a) can be inspected more closely in figure \ref{fig_psd}. In panel a), the connected lipid phase is depicted as a labelmap representation from the applied watershed algorithm where each lipid domain is represented in a different color. Visually, the lipid network seems to consist of spherical structures aggregated into a close packing. This is consistent with a partial coalescence of the fat globules. Although some labels are clearly non-spherical in shape, the algorithm has identified a large number of spherical lipid domains that resemble individual fat globules. The resulting size distributions of the equivalent diameters of all four samples are shown in panel b). It is seen that the samples give almost identical size distributions with the mode at around $1 \umum$. The mean values of the equivalent diameters are between $1.2 - 1.4 \umum$, as seen in table \ref{tab_eden}. These distributions closely resemble previously reported distributions of the fat globule sizes using dynamic light scattering \citep{munk_whippable_2013,munk_coalescence_2015}. Since these latter were measured on control PKO emulsions without added LACTEM or GMU, the present findings in panel b) support the hypothesis of partial coalescence.


\begin{table}[tb]
\centering
\renewcommand{\tabcolsep}{2.5mm}
\renewcommand{\arraystretch}{1.2}
\begin{tabular}{l|c|c|c|c|c}
\multicolumn{6}{c}{ \textbf{Percent object volume (POV)}}\\
\hline
Data type             & Ref & Sample 1 & Sample 2 & Sample 3 & Sample 4 \\
		      & [vol\%]    & [vol\%]& [vol\%]& [vol\%]& [vol\%]\\
\hline
Lipid phase           & 28.4       &  31.2  & 38.1   & 37.7   & 26.5 \\
Water phase           & 71.4       &  68.7  & 61.9   & 62.2   & 73.5 \\
Cellulose             &  0.2       &  0.05  &  0.01  &  0.03  & 0.02\\
\hline
\multicolumn{6}{c}{ \textbf{Electron densities, $\rho_e$}}\\
\hline
		      & [$\textup{e}/\angstrom^3$]  & [$\textup{e}/\angstrom^3$] & 
		      [$\textup{e}/\angstrom^3$] & [$\textup{e}/\angstrom^3$] & [$\textup{e}/\angstrom^3$] \\
\hline
Lipid phase           & 0.308 & 0.314 & 0.323 & 0.321 & 0.318 \\
Water phase           & 0.348 & 0.362 & 0.363 & 0.360 & 0.360 \\
Cellulose             & 0.477 & 0.397 & 0.382 & 0.391 & 0.374 \\
Micropipette          & 0.666 & 0.668 & 0.671 & 0.673 & 0.668 \\
\hline
Bulk                  & 0.336 & 0.347 & 0.348 & 0.345 & 0.349 \\
\hline
\multicolumn{6}{c}{ \textbf{Mean equivalent diameter, $d$}}\\
\hline
		      & [$\umu \textrm{m}$] & [$\umu \textrm{m}$]& [$\umu \textrm{m}$]& [$\umu \textrm{m}$]& [$\umu \textrm{m}$]\\
\hline
Lipid phase           & 0.98  & 1.15  & 1.41  & 1.40  & 1.25 \\
\hline 
\end{tabular}
\caption{\label{tab_eden}Percent object volume (POV), mean electron densities and equivalent lipid diameter from the four tomograms compared to reference values. The bulk emulsion electron density was calculated from the $\rho_e$ and POV values of the emulsion phases in the table. Electron density values for the micropipette are included as well.}
\end{table}

Measured and reference POV values are shown in table \ref{tab_eden}. POV of the cellulose was lower than the expected in all tomograms. This might be due to the confined space in the micropipette which would not permit larger pieces of microcrystalline cellulose. The POV for lipid- and water phase varied between the tomograms with a larger lipid phase POV in sample 1-3 than expected and a smaller in sample 4. These variations demonstrate the challenge of obtaining a representative sampling with volumes on the micron scale. Although sampled within $100\umum$ on the same micropipette, large variations in the POV values of sample 3 and 4 were observed. 

In addition to the spatial information, PXCT also provides information on the electron density distribution in the sample. In table \ref{tab_eden}, mean electron densities for the emulsion phases and the micropipette phase are shown for all measured samples as well as reference values. For all phases, the measured mean electron density values are quite similar across the samples. In addition, the precision of the PXCT measurements can be estimated by comparing the measured micropitte values with the reference values. For all samples, the obtained mean electron density for the micropipette phase was within 1\% of the reference value as seen in table \ref{tab_eden}.
A bit worse agreement was found when comparing measured and reference values of the emulsion phases in table \ref{tab_eden}. For the lipid and water phases, the measured mean electron densities were around 4\% higher than the reference values. On the other hand, the cellulose from the stabilizer mixture displayed values 20\% lower than the reference. For the cellulose, the size of the individual cellulose pieces was small compared to the spatial resolution which would result in a lower observed mean value.

\section{Discussion}

The main focus in the present study regards the extend, structure and composition of the lipid phase in the PKO emulsion.  For all samples, the PXCT measurements revealed an extended network connecting almost all of the fat globules. In addition, the morphology of the network was consistent with a partial coalescence of the fat globules to an extreme degree. These findings were in accordance with previous studies \citep{munk_whippable_2013,munk_coalescence_2015}. No clear structural differences between the two formulations with LACTEM and LACTEM+GMU were apparent. Previously, the two formulations have also been found to be similar in both microstructure and elastic modulus \citep{munk_whippable_2013}. While a single- or two-phase system is often employed when studying extended networks in food systems, the sample PKO emulsion used here constitutes a full multiphase food system. This indicates that PXCT can be used to investigate even complex food systems.

When considering the small sample volume in the PXCT measurements, a relevant question is how these findings relate to the bulk emulsion. As the micropipettes used for PXCT confined the PKO emulsions to a few tens of micrometers in diameter, the observed structure might be different than for bulk emulsion. This could be due to an applied shear stress during injection or a change in fat globule size distribution due to the micropipette width. However, while the amount and size of microcrystalline cellulose was reduced, the lipid morphology and distributions of domain diameters were in agreement with previous findings for bulk PKO emulsion. In addition, no air pockets larger than the spatial resolution were observed indicating an absence of large shear stresses. Furthermore, a presence of sub-resolution air pockets would result in a reduction of the observed electron density. Thus, since the measured electron densities for the lipid and water phases were higher than the reference values, a large presence of sub-resolution air pockets is unlikely.

Since PXCT provides quantitative electron density values, these values may be used to identify the known phases in the sample and highlight unknown or unexpected. Hence, since the observed mean electron densities for the water and lipid phases were within 4\% of the reference values, these phases could be easily recognized in the reconstructed tomograms. In addition, the quantitative electron densities allowed to distinguish the stabilizer phase even though the POV was below 1\%.

When applying PXCT for food products, the sample stability becomes an issue. As described above, radiation damage due to the initial high X-ray flux required a reduction in the beam flux. However, as this reduction prolonged the total scan time, movement of the sample during measurements became an issue for the spatial resolution. 
A way to avoid these challenges would be to measure the sample in a frozen state. Hence both stability and resistence towards radiation damage would be increased. However, this will not be an option for studies of the native state at room temperature as in the present study.

Compared to other X-ray phase-contrast imaging techniques, a main advantage of PXCT is the submicron spatial resolution offered. In comparison, X-ray grating interferometry is limited to the micron range by the period of the gratings used. Propagation-based techniques for X-ray phase-contrast imaging are restricted by the (effective) pixel size of the detection system, and are typically also in the micron-range resolution. 
An X-ray phase-contrast technique which has achieved sub-micron resolution for biological samples is full-field tomographic microscopy with Zernike phase-contrast \citep{stampanoni_nanoscale_2010}. Potentially, this could be an alternative way to resolve the structures of the PKO emulsion. However, quantitative reconstruction of the electron density is challenging using this technique compared to PXCT.

Compared with CLSM, PXCT offers a more isotropic PSF which results in a uniform spatial resolution in both horizontal and vertical direction. Furthermore, PXCT is not limited to investigations just below the surface of the sample. In addition, PXCT measures the unaltered sample without staining as in CLSM, and hence provides the electron densities of all phases present. Thus, besides the structure of the lipid phase, also the water phase, microcrystalline cellulose and the absence of air pockets were observed in PXCT.

One of the limitations of PXCT is that the horizontal FoV is restricted to a few tens of micrometers. Also, while e.g. CLSM can be performed using tabletop lap equipment, PXCT is limited to synchrotron facilities.

\section{Conclusion}

This study has shown the feasibility of PXCT for food science applications. The obtained 3D microstructure of the PKO emulsion constitutes the first reported direct measurement of an extended network of fat globules in a food product. In the PKO emulsion. the lipid network was structed by aggregated lipid domains and almost the entire lipid fraction was included in the extended network. This demonstrated an extreme case of partial coalescence of fat globules in a oil-in-water PKO emulsion. The observed sizes and shapes of the lipid domains were in agreement with CLSM micrographs and previous measurements of fat globule size distribution.

The electron density values measured by PXCT were consistent across all measured samples. For the micropipette, the measured mean values were within 1\% of the reference value. For the lipid and water phases, the values were within 4\% of the expected value. Thus, as a quantitative method, PXCT can be used to investigate the composition of the sample. In the present study, this was used to rule out the presence of sub-resolution air pockets.

In further work, the 3D network structures obtained from the PXCT measurements could be used for fractal analysis as in e.g. \citep{narine_physreve_1999,marangoni_edible_2012}. From such analysis, a link between the microstructure and macroscopic mechanical properties of the emulsion could be established. 

Altogether, PXCT is a promising tool for spatial and quantitative investigation of food products on the sub-micron scale. This gives a novel quantitative experimental approach for direct studies of the 3D microstructure of food materials. 
In order to achieve higher spatial resolution, considerations regarding sample stability and radiation damage must be taken into account.

\section*{Acknowledgements}
      
The authors would like to acknowledge the assistance of Kristian Rix in the PXCT measurements, and the assistance of Poul Martin Bendix with equipment and guidance to prepare the micropipettes. The authors are grateful for the fruitful discussions with Pil Maria Saugmann on the manuscript.
The study received funding through the Danish Strategic Research Council through the NEXIM project.



\bibliography{/home/schou/Documents/NEXIM_and_xray/artikelproduktion/ptycho_emulsions/ptycho_emulsion_papers2.bib}

      \end{document}